\documentclass[aps,pra,twocolumn,floats,showpacs]{revtex4}
\usepackage{epsfig}
\usepackage{color}
\usepackage{bm}
\usepackage{latexsym}

\begin{document}
\newcommand{\hide}[1]{}
\newcommand{\tbox}[1]{\mbox{\tiny #1}}
\newcommand{\half}{\mbox{\small $\frac{1}{2}$}}
\newcommand{\sinc}{\mbox{sinc}}
\newcommand{\const}{\mbox{const}}
\newcommand{\trc}{\mbox{trace}}
\newcommand{\intt}{\int\!\!\!\!\int }
\newcommand{\ointt}{\int\!\!\!\!\int\!\!\!\!\!\circ\ }
\newcommand{\eexp}{\mbox{e}^}
\newcommand{\bra}{\left\langle}
\newcommand{\ket}{\right\rangle}
\newcommand{\EPS} {\mbox{\LARGE $\epsilon$}}
\newcommand{\ar}{\mathsf r}
\newcommand{\im}{\mbox{Im}}
\newcommand{\re}{\mbox{Re}}
\newcommand{\bmsf}[1]{\bm{\mathsf{#1}}}
\newcommand{\mpg}[2][1.0\hsize]{\begin{minipage}[b]{#1}{#2}\end{minipage}}

\title{Photonic heterostructures with L\'evy-type disorder: statistics of coherent transmission}

\author{A. A. Fern\'andez-Mar\'in and J. A. M\'endez-Berm\'udez}

\affiliation{Instituto de F\'{\i}sica, Benem\'erita Universidad Aut\'onoma de Puebla,
Apartado Postal J-48, Puebla 72570, Mexico}

\author{Victor A. Gopar}
\affiliation{Departamento de F\'isica Te\'orica, Facultad de Ciencias, and Instituto de Biocomputaci\'on y F\'isica de Sistemas Complejos,  Universidad de Zaragoza, Pedro Cerbuna 12, E-50009 Zaragoza, Spain}

\date{\today}

\begin{abstract}
We study the electromagnetic transmission $T$ through one-dimensional (1D)
photonic heterostructures whose random layer thicknesses follow a long-tailed distribution --L\'evy-type distribution. Based on recent predictions made for 1D coherent  transport with L\'evy-type disorder, we show numerically  that for a system of length $L$
(i) the average $\left\langle - \ln T \right\rangle \propto L^\alpha$ for $0<\alpha<1$,
while $\left\langle - \ln T \right\rangle \propto L$ for $1\le\alpha<2$, $\alpha$ being the exponent of the power-law decay of
the layer-thickness  probability distribution; and
(ii) the transmission distribution $P(T)$ is independent of the angle of incidence  and frequency of the electromagnetic wave,  but  it  is fully determined by the values of $\alpha$ and  $\left\langle \ln T \right\rangle$. Additionally we have found and numerically verified  that $\left\langle T \right\rangle \propto L^{-\alpha}$ with $0 < \alpha <1$.
\end{abstract}


\maketitle

Random processes characterized by density probabilities with a long tail (L\'evy-type processes)  have been found and studied
in very different phenomena and fields such as biology, economy, and physics. One of the main features of
a L\'evy-type density distribution $p(l)$ is the slow decay of its tail. More precisely, for large $l$,
\begin{equation}
p(l) \sim \frac{1}{l^{1+\alpha}} \ ,
\label{levy}
\end{equation}
with $0 < \alpha <2$. Note that the second moment diverges for all $\alpha$
and if $0< \alpha <1$ also the first moment diverge. This kind of distributions
are also known as $\alpha$-stable distributions \cite{uchaikin}.
A window on new optical materials which allow for the experimental study of L\'evy flights in an outstanding controllable way was recently opened
with the construction of the so-called {\it L\'evy glass} \cite{BBW08}:
titanium dioxide particles are suspended in a matrix made of glass microspheres. The distribution
of the microsphere diameters is properly chosen in order that light can travel performing  L\'evy
flights within the microspheres. The diameter distribution is characterized by the exponent $\alpha$ of the power-law decay
of its tail; it was found \cite{BBW08} that when $0<\alpha<1$ the transport is supperdiffusive, while for $\alpha=2$
the normal diffusive transport is recovered. This experimental investigation has motivated several theoretical works on the effects of
the presence of L\'evy-type processes on
different transport quantities in one dimension, as well as in higher dimensional systems \cite{beenakker,burioni,buosante,FG10, sibatov}.

On the other hand, coherent electron transport through one-dimensional (1D)
quantum wires with L\'evy-type disorder was studied in
Ref.~\cite{FG10}. It was found that for the dimensionless conductance, or transmission,
$T$:
\begin{itemize}
\item[{\bf (i)}]
the average (over different disorder realizations) of the logarithm of the transmission  behaves as
\begin{eqnarray}
\left\langle  - \ln T \right\rangle \propto \left\{\begin{array}{ll}
L^\alpha & \mbox{for} \quad 0 < \alpha < 1  \\
L        & \mbox{for} \quad 1 \le \alpha < 2
\end{array}\right. \ ,
\label{avlnT}
\end{eqnarray}
and
\item[{\bf (ii)}] the distribution of transmission $P(T)$
is fully determined by the exponent $\alpha$ and the ensemble average
$\left\langle \ln T \right\rangle$.
\end{itemize}
We point out that although for $1 \le \alpha < 2$, the average $\left\langle \ln T \right\rangle$
depends linearly on $L$, as in the standard Anderson localization problem, it is
interesting to remark that the statistical properties of $T$ are not those predicted
by the standard scaling approach to localization, in particular by the
Dorokhov-Mello-Pereyra-Kumar (DMPK) equation \cite{MK04}. That is, for
$1 \le \alpha < 2$ the transmission fluctuations are larger than those considered
in the DMPK equation. Thus, the standard statistical properties of $T$ are recovered for $\alpha \ge 2$.

Having as a reference various analogies between electron, light, and
matter-wave transport \cite{var0, var1,var2,var3,var4} one may expect the statements
{\bf (i)} and {\bf (ii)} to be also valid for 1D optical systems,
even though in the latter case additional parameters such as incidence
angle and frequency come into play. A systematic investigation of the statistical
properties of coherent transmission for a 1D analog of
L\'evy glasses is not available in the literature and the above expectations
have not been verified until now.
Therefore, in this paper
we undertake this task by studying the transmission $T$ through 1D photonic
heterostructures with L\'evy-type spacing disorder.
Moreover, engineering the disorder in a photonic
heterostructure might be a less complex task than in an electronic system; so we
hope that this work stimulate future  photonic experiments.

\begin{figure}[t]
\centerline{\includegraphics[width=8cm]{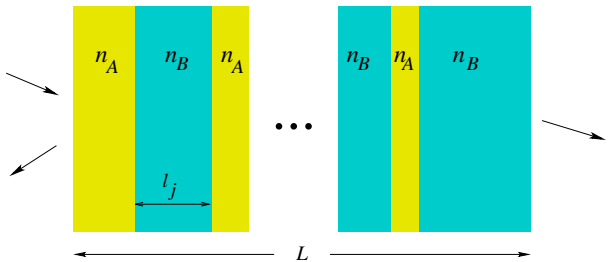}}
\caption{(Color online) Sketch of a heterostructure of length $L$
with random layer thicknesses $l_j$ and refractive indexes $n_{A(B)}$.
The distribution of the thicknesses follows a L\'evy-type distribution.}
\label{Fig1}
\end{figure}

The heterostructures that we shall study consist of an alternating sequence
of layers of materials $A$ and $B$ having refractive indexes $n_A$ and
$n_B$, respectively. See Fig.~\ref{Fig1}. The corresponding thicknesses
$l_A$ and $l_B$ are chosen as random numbers drawn from a L\'evy-type  distribution characterized by the exponent $\alpha$ of its power-law tail.
In this work we only consider L\'evy-type distributions supported along
the positive semiaxis and  focus on the case $0< \alpha <1$, where large fluctuations of the transmission produce the most interesting effects. The total length $L$ of the heterostructure is then given by $L=\sum l_j$.
Without lost of generality we consider that the heterostructure
grows in the $z$-direction.

\begin{figure*}[t]
\includegraphics[width=9cm]{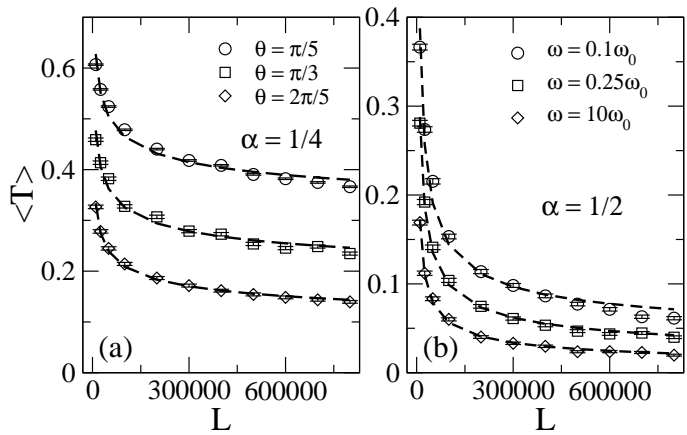}\hspace{2pc}
\begin{minipage}[b]{16pc}
\caption{Average transmission $\left\langle T \right\rangle$ as a function
of $L$ (symbols) for 1D heterostructures with L\'evy-type spacing disorder
characterized by $\alpha$. In (a) [(b)] three values of $\theta$ [$\omega$]
were considered for $\omega=0.25$ [$\theta=0$]. Dashed curves are fittings
of the numerical data accordingly to Eq.~(\ref{avT1}). Each symbol was calculated using $10^4$ ensemble realizations.}
\label{Fig2}
\end{minipage}
\end{figure*}

1D heterostructures with L\'evy-type thickness  disorder, as defined above, can
be produced with porous silicon. Layers of silicon with different porosities
(i.e. different refractive indexes) and varying thicknesses can be obtained by HF
electrochemical etching by modulating the value of the current density during
the anodization process \cite{BLRG03}.
Thus we fix $n_A=1.4$ and $n_B=2.4$ in our calculations since they
correspond to experimentally accessible porous silicon refractive index values
\cite{BLRG03,EAMA08}.

We compute the transmission $T$ through our 1D heterostructure using the transfer
matrix formalism described in Ref.~\cite{MS08}. We consider an electromagnetic
wave with frequency $\omega$ that strikes the first layer of the
heterostructure (embedded in air) making an angle $\theta$ with respect
to the $z$-axis, then propagates inside the structure composed of $N$
layers, and finally escapes through the layer on the opposite side.
Without loss of generality, in the following we specialize on TE modes.
The transfer matrix of the scattering process can be written
as \cite{MS08}
\begin{equation}
{\bf M}={\bf M}_{0,N}{\bf M}_{\mbox{\tiny het}}{\bf M}_{1,0} \ ,
\end{equation}
where ${\bf M}_{\mbox{\tiny het}} = {\bf D}_{N}{\bf M}_{N,N-1}{\bf D}_{N-1}\cdots {\bf M}_{2,1}{\bf D}_{1}$,
\begin{eqnarray}
{\bf D}_{j} & = & \left(\begin{array}{cc}
\exp(ik_{jz}l_j) & 0 \\
0 & \exp(-ik_{jz}l_j) \\
\end{array}
\right)  \ , \nonumber \\
{\bf M}_{j,j-1} & = & \frac{1}{2}\left(\begin{array}{cc}
1+k_{(j-1)z}/k_{jz} & 1-k_{(j-1)z}/k_{jz} \\
1-k_{(j-1)z}/k_{jz} & 1+k_{(j-1)z}/k_{jz} \\
\end{array}
\right) \ , \nonumber
\end{eqnarray}
and
$k_{jz}$ is the component of the wave vector along the $z-$direction
in the $j$th layer given by $k_{jz}=k_j\cos(\theta_j)$ with
$k_j=(\omega/c)n_j$ ($n_j$ equals $n_A$ or $n_B$).
$\theta_j$ is related to $\theta_{j-1}$ through Snell's law:
$\sin(\theta_j)/\sin(\theta_{j-1})=n_{j-1}/n_j$.
Above,
${\bf M}_{1,0}$ [${\bf M}_{0,N}$] is the transfer matrix for
the interface between the first [$N$th] layer and  air, while
${\bf M}_{\mbox{\tiny het}}$ is the transfer matrix of the heterostructure.
Finally,
\begin{equation}
T=\left| {\bf M}_{22} \right|^{-2} \ .
\end{equation}
In the following we shall study the statistical properties of $T$, in particular its full distribution. Concerning the numerical simulations along this work, the statistics is collected over an ensemble of different disorder realizations. Notice that since the thicknesses $l_i$ are drawn from a L\'evy-type  distribution, Eq.~(\ref{levy}), for a fixed length $L$ the number of layers composing a heterostructure might vary strongly from sample to sample.
Along the paper, $\omega$ is given in units
of the reference frequency $\omega_0=2\pi c/\lambda_0$ where $\lambda_0$
could be chosen to provide suitable experimental conditions.

As we have mentioned, the theoretical results in this paper  are based on the analysis presented in \cite{FG10}. Here we only  reproduce the main result of that work:  the distribution for transmission $P_\xi (T)$ with $\xi \equiv  \langle \ln T \rangle$ . The average quantities that we study, such as $\langle T \rangle$ and $\langle \ln T \rangle$ [Eq.~(\ref{avlnT})] are derived from the following expression for the distribution
\begin{eqnarray}
\label{pofG_xi}
P_\xi(T)=\int_0^\infty p_{s(\alpha,\xi,x)}(T) q_{\alpha,1}(x){\rm d}x \ ,
\end{eqnarray}
for $0< \alpha <1$, where  $q_{\alpha,c}$ is the probability density function of the L\'evy-type distribution supported in the positive semiaxis,
$
s(\alpha,\xi,x)={\xi}/(2{x^\alpha I_\alpha)}
$,
$
I_\alpha =1/2 \int_{0}^{\infty} x^{-\alpha} q_{\alpha,1}{\rm d}x
$, and
\begin{equation}
\label{pofg}
p_s(T)=\frac{s^{-3/2}}{\sqrt{2\pi}}
\frac{\exp(-s/4)}{T^2}\int_{y_0}^{\infty}dy\frac{y \ \exp(-y^2/4s)}
{\sqrt{\cosh{y}+1-2/T}} \ ,
\end{equation}
where $y_0={\rm arcosh}{(2/T-1)}$.

\begin{figure*}[t]
\includegraphics[width=9cm]{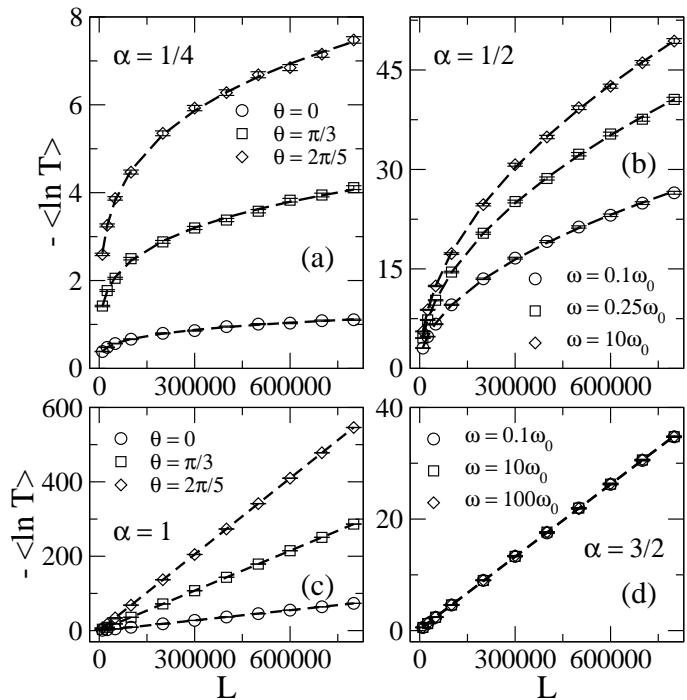}\hspace{2pc}
\begin{minipage}[b]{16pc}
\caption{The average $-\left\langle \ln T \right\rangle$ as a function of
$L$ (symbols) for 1D heterostructures with L\'evy-type spacing disorder
characterized by $\alpha$. In (a,c) [(b,d)] three values of $\theta$
[$\omega$] were considered for $\omega=0.25$ [$\theta=0$].
The dashed curves are fittings of the data with Eq.~(\ref{avlnT}).
Each symbol was calculated using $10^4$ ensemble realizations.}
\label{Fig3}
\end{minipage}
\end{figure*}

We start by analyzing the average transmission
$\left\langle T \right\rangle$. From Eq.~(\ref{pofG_xi}), we find that
\begin{equation}
\left\langle T \right\rangle \propto L^{-\alpha} \ ,
\label{avT1}
\end{equation}
for $0< \alpha < 1$. We have verified numerically this result. In Fig.~\ref{Fig2} we present the average transmission for different values of $\theta$ and $\omega$ for $\alpha = 1/4$ and $\alpha = 1/2$. As we can see, the agreement is very good in all cases.
This result, Eq.~(\ref{avT1}),  could be contrasted with the exponential decay of $\langle T \rangle$ with $L$,   for standard 1D disordered systems, and with $\left\langle T \right\rangle \propto 1/L$,  for quasi 1D systems in the  normal diffusive transport regime.

We now study the average $\left\langle \ln T \right\rangle$. In
electronic transport this quantity is of relevance since it gives information about
the localization length of the disordered system. In Fig.~\ref{Fig3} we present
different plots of the average
$\left\langle \ln T \right\rangle$ as a function of $L$ for $\alpha<1$
[Fig.~\ref{Fig3}(a-b)] and $\alpha\ge 1$ [Fig.~\ref{Fig3}(c-d)].
For $\alpha<1$ we observe a clear behavior of the form
$\left\langle - \ln T \right\rangle \propto L^{\alpha}$; while for
$\alpha\ge 1$ we see that $\left\langle - \ln T \right\rangle$ is simply
proportional to $L$, as for standard 1D disordered systems. Therefore,
all curves in Fig.~\ref{Fig3} are  well
described by Eq.~(\ref{avlnT}).
In addition, notice that while the curves of $\left\langle \ln T \right\rangle$
{\it vs} $L$ strongly depend on the incidence angle $\theta$ for all $\alpha$, see
Fig.~\ref{Fig3}(a) and \ref{Fig3}(c), the dependence on $\omega$ is lost for
$\alpha\ge 1$, see Fig.~\ref{Fig3}(b) and \ref{Fig3}(d).

\begin{figure*}[t]
\includegraphics[width=9cm]{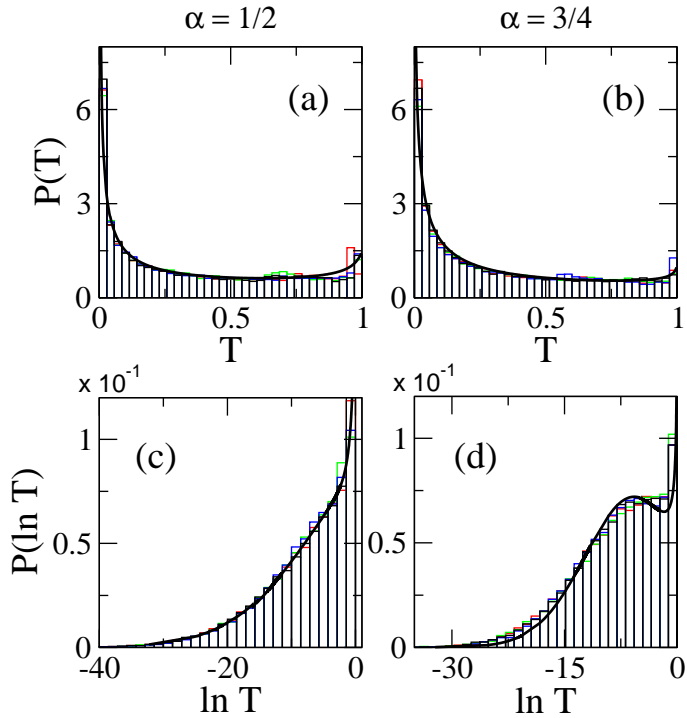}\hspace{2pc}
\begin{minipage}[b]{16pc}
\caption{(Color online) The probability distribution functions $P(T)$
(a-b) and $P(\ln T)$ (c-d) for $\alpha=1/2$ and 3/4 (histograms).
Each panel contains 4 histograms with values of $\theta$ and $\omega$
given in Table~\ref{Table1}.
The histograms in (a-b) [(c-d)] are characterized by
$-\left\langle \ln T \right\rangle \approx 2$
[$-\left\langle \ln T \right\rangle \approx 8$].
Black dashed curves are the corresponding theoretical predictions
obtained from Eq.~(\ref{pofG_xi}).
Each histogram was calculated using $3\times10^4$ ensemble realizations.}
\label{Fig4}
\end{minipage}
\end{figure*}
\begin{table}[t]
  \centering
  \begin{tabular}{ccccc} \\ \hline
    panel & RED  & BLUE & GREEN & BLACK \\ \hline
(a) & (0,0.1)  & ($\pi/5$,0.5) & ($\pi/4$,10) & ($3\pi/7$,1) \\
(b) & (0,10) & ($\pi/5$,1) & ($\pi/6$,0.1) & ($\pi/3$,0.5) \\
(c) & (0,10)  & ($\pi/5$,5) & ($2\pi/5$,1) & ($\pi/6$,0.1) \\
(d) & (0,2) & ($\pi/4$,10) & ($\pi/5$,1) & ($\pi/3$,5) \\
  \hline
  \end{tabular}
\caption{Values of $(\theta,\omega/\omega_0)$ used for the histograms (labeled by their color) in Fig.~\ref{Fig4}.}
\label{Table1}
\end{table}

Next, we analyze the full distribution of the transmission.
It is clear from Fig.~\ref{Fig3} that by choosing the appropriate
combination of $\theta$, $\omega$ and $L$ one can fix the value of
the average $\left\langle \ln T \right\rangle$ for a given $\alpha$.
Then, in Fig.~\ref{Fig4}(a-b) we show probability distribution
functions $P(T)$ for $\alpha=1/2$ and 3/4 for
$\left\langle - \ln T \right\rangle =2$. Since for
smaller values of $\left\langle \ln T \right\rangle$, $P(T)$
is concentrated close to $T=0$, in Fig.~\ref{Fig4}(c-d) we present
$P(\ln T)$ for $\left\langle - \ln T \right\rangle \approx 8$.
Notice that each panel in this figure contains 4 histograms with different
combinations of $\theta$ and $\omega$. However, all histograms fall one on top
of the other; that is, $P(T)$ and $P(\ln T)$ are completely  determined by $\alpha$
and $\left\langle \ln T \right\rangle$.
Moreover, the black dashed lines are the corresponding theoretical predictions for $P(T)$
and $P(\ln T)$ for the specific
combinations of $\alpha$ and $\left\langle \ln T \right\rangle$ we used.
Evidently, the correspondence between theory and numerics is excellent affirming
the equivalence between quantum and electromagnetic 1D disordered systems.

Finally, we want to stress that even though we have used arbitrary units for the
length $L$ of the heterostructures, our results may be experimentally confirmed
by properly choosing the frequency $\omega=2\pi c/\lambda$.
For instance, since $\omega/\omega_0 = \lambda_0/\lambda$, if
$\lambda_0=2L$
a typical experiment in the visible range with $\lambda \sim 500\mbox{nm}$
and $L\sim 10\mu$ \cite{BLRG03,EAMA08} will set the ratio $\omega/\omega_0$
to 40. Then, $\theta$ can be tuned to specify a desired value
of $\left\langle T \right\rangle$ or $\left\langle \ln T \right\rangle$.

In summary, for 1D heterostructures whose layer thicknesses follow a L\'evy-type
distribution, characterized by a  power-law decay,
we have shown that: $\left\langle - \ln T \right\rangle \propto L^\alpha$ for
$0< \alpha <1$; and once $\alpha$ and
$\left\langle \ln T \right\rangle$ are fixed the distribution of transmission
$P(T)$ is invariant with respect to the system parameters ($\theta, \omega$).
Also, we have found that $\left\langle T \right\rangle \propto L^{-\alpha}$, in contrast to
the exponential decay in standard disordered 1D systems.
We have verified that our results are unaffected by considering TM modes and
different refraction index contrasts $n_A/n_B$.

We hope that  our results  serve to deepen our understanding of transport
properties when L\'evy-type disorder is present and motivate further experimental
investigation on its unconventional effects, as those shown in this work.

\begin{acknowledgments}
This work was partially supported by
VIEP-BUAP (project MEBJ-EXC10-I), Mexico, and
MICINN (project FIS2009-07277), Spain.
\end{acknowledgments}

\end{document}